# Combining Search, Social Media, and Traditional Data Sources to Improve Influenza Surveillance.

**Authors:** M. Santillana[1,2,3], A. T. Nguyen[1], M. Dredze[4], M. J. Paul[5], and J. S. Brownstein[2,3]

**Affiliations:**

[1] Harvard School of Engineering and Applied Sciences, Cambridge, MA

[2] Boston Children's Hospital Informatics Program, Boston, MA

[3] Harvard Medical School, Boston, MA

[4] Department of Computer Science, Johns Hopkins University, Baltimore, MD

[5] Department of Information Science, University of Colorado, Boulder, CO

**Scientific Abstract**

We present a machine learning-based methodology capable of providing real-time ("nowcast") and forecast estimates of influenza activity in the US by leveraging data from multiple data sources including: Google searches, Twitter microblogs, nearly real-time hospital visit records, and data from a participatory surveillance system. Our main contribution consists of combining multiple influenza-like illnesses (ILI) activity estimates, generated independently with each data source, into a single prediction of ILI utilizing machine learning ensemble approaches. Our methodology exploits the information in each data source and produces accurate weekly ILI predictions for up to four weeks ahead of the release of CDC's ILI reports. We evaluate the predictive ability of our ensemble approach during the 2013-2014 (retrospective) and 2014-2015 (live) flu seasons for each of the four weekly time horizons. Our ensemble approach demonstrates several advantages: (1) our ensemble method's predictions outperform every prediction using each data source independently, (2) our methodology can produce predictions one week ahead of GFT's real-time estimates with comparable accuracy, and (3) our two and three week forecast estimates have comparable accuracy to real-time predictions using an autoregressive model. Moreover, our results show that considerable insight is gained from incorporating disparate data streams, in the form of social media and crowd sourced data, into influenza predictions in all time horizons.

**Author Summary**

The aggregated activity patterns of Internet users have enabled the detection and tracking of multiple population-wide events such as disease outbreaks, financial markets performance, and preferences in online movie selections. As a consequence, a collection of mathematical models aiming at monitoring and predicting these events in real-time have been proposed in the past decade. As we discover new methods and data sources suitable to track these events, it is not clear whether more information will lead to improved predictions. In the context of digital disease detection at the population level, we show that it is advantageous to combine the information from multiple flu activity predictors in the US than simply choosing the best performing flu predictor. Our findings suggest that the information from multiple data sources such as Google searches, Twitter microblogs, nearly real-time hospital visit records, and data from a participatory surveillance system, complement one another and produce the most accurate and robust set of flu predictions when combined optimally.

## 1. Introduction

Predicting the dynamics of seasonal and non-seasonal influenza outbreaks remains a great challenge [1]. They cause up to 500,000 deaths a year worldwide and an estimated 3,000 to 50,000 deaths a year in the United States of America (US) [2]. Frequently, their severity cannot be assessed in a timely manner, and thus, systems capable of providing estimates of influenza incidence are critical to allow health officials to properly prepare for and respond to influenza-like illness (ILI) outbreaks. The US Centers for Disease Control and Prevention (CDC) continuously monitor the level of ILI circulation in the US population by gathering information from physicians' reports that record the percentage of patients seen in clinics who exhibit influenza-like illnesses (ILI) symptoms. While CDC ILI data provides public health officials with an important proxy of influenza activity in the population, its availability has a known lag-time of at least 7 to 14 days. This means that by the time the data is available, the information is already 1 or 2 weeks old.

Many attempts have been made to estimate the ILI activity in the US ahead of the release of CDC reports, some using a combination of statistical and mechanistic SIR models [3,4,5] and others using non-traditional Internet-based information systems such as: Google [6,7], Yahoo [8], and Baidu [9] Internet searches, Twitter posts [10,11,12], Wikipedia article views [13,14], Flu Near You [15,16], and clinicians' databases (such as UpToDate) queries [17]. We will focus on non-traditional Internet-based approaches here. Google Flu Trends (GFT) [6], a widely accepted digital disease detection system that uses the Google search volume of specific terms to predict ILI in the US and other countries, continuously provides real-time estimates of ILI. Even though GFT was initially hailed as a success, its inaccuracies in multiple time periods of high ILI have lead to doubts about the utility of these data [18]. While Google and external researchers have worked to update and reevaluate the methodology behind GFT [19, 20, 21, 22, 23, 24], alternative and independent methods to estimate ILI in real-time are still needed.

We propose a methodology based on machine learning algorithms capable of providing real-time ("nowcast") and forecast estimates of ILI by leveraging data from multiple sources including: Google searches, nearly real-time hospital visit records provided by athenahealth, Twitter posts, and data from Flu Near You, a participatory surveillance system. While models using these data sources to predict ILI may capture different flu incidence signals in the population, we show that they complement one another when we combine them to predict CDC's ILI. Our main contribution consists of optimally combining multiple ILI estimates, generated independently with each data source, into a single prediction of ILI utilizing machine learning ensemble approaches. Our methodology exploits the information in each data source and produces accurate weekly ILI predictions for up to four weeks ahead of the release of CDC's ILI reports, effectively producing forecasts three weeks into the future. We evaluate the predictive ability of our ensemble approach during the 2013-2014 and 2014-2015 flu seasons for each of the four weekly time horizons.

## 2. Data

We collected CDC-reported ILI, considered the ground truth for this study, from the ILINet website (http://gis.cdc.gov/grasp/fluview/fluportaldashboard.html). We used five independent data sets to develop our ILI weak predictors: (a) near real-time hospital visit records from athenahealth, a medical practices management company; (b) Google Trends, a Google service that provides approximate search volumes for specific queries (www.google.com/trends), (c) influenza-related Twitter microblogging posts, (d) FluNearYou, a participatory surveillance system to self-report ILI; and (e) Google Flu Trends. All datasets were accessed and downloaded on March 16, 2015.

*CDC Data*

The CDC compiles data on the weekly number of people seeking medical attention with ILI symptoms in the United States. CDC's ILI data is freely distributed and available through ILInet, via the online FluView tool, which posts both new and historical data (http://gis.cdc.gov/grasp/fluview/fluportaldashboard.html). Typically, new CDC reports provide a first estimate of %ILI and as more reports are received, revised CDC reports are released and become the official %ILI. We used the revised CDC reports for weeks 1/10/04 to 02/21/15 as our gold standard for validation purposes. For training our models, we used the (then available) unrevised CDC reports. Weekly tables released on week **W** of the **X-Y** season are available at the following URLs: http://www.cdc.gov/flu/weekly/weeklyarchives**X-Y**/data/senAllregt**W**.htm

See [11] for details on obtaining and using historical CDC data.

Additionally, the CDC reports the number and percentage of laboratory tests that are positive for influenza types A and B, using data reported by WHO and NREVSS collaborating laboratories across the United States. This virology data is not part of the ensemble but is used for comparison. Similar to the ILI data, the virology data is subject to weekly revisions, which can be obtained through weekly tables available at: http://www.cdc.gov/flu/weekly/weeklyarchives**X**-**Y**/data/whoAllregt**W**.html

*athenahealth Data*

We obtained weekly nationally aggregated data reporting the total number of people seeking medical attention with ILI symptoms in medical practices managed by athenahealth, from Jul 2009 to Feb 2015. athenahealth data is typically available at least one week ahead of CDC ILI reports. By dynamically finding the best linear model to historically map athenahealth's ILI onto CDC's ILI, we were able to produce (out-of-sample) ILI estimates using athenahealth's data as a predictor, one week ahead of CDC reports during our study period. We refer to this data as ATH in the plots and tables. We used ATH data for weeks 6/28/09 to 02/21/15.

*GT Data*

Following the methodology proposed in [17] and [22], we used data from Google Trends (GT) as a proxy of the volume of query searches for 100 search terms and then utilized a dynamic multivariate approach to predict flu activity for the time period Jul 2013 - Feb 2015. The *logit* transform utilized in [22] was not used to produce our out-of-sample predictions since the identity transformation [17] showed better performance. We used GT data for weeks 1/10/04 to 02/21/15.

*Twitter Data*

We used the Twitter (TWT) influenza classification system introduced by [25,26], which identifies Twitter messages that express an influenza infection. The logistic regression classifiers were trained on approximately 12,000 tweets annotated for relevance, distinguishing tweets that indicated an infection rather than discussing influenza in other contexts. The normalized weekly volumes of influenza tweets are available from HealthTweets.org [27]. ILI predictions are then created by including the influenza tweet volumes in a linear autoregression exogenous (ARX) model, as described in [6], using the previous three weeks of CDC-reported ILI. The Twitter data spans 11/27/11 – 2/15/15 and the CDC data (starting three weeks before Twitter) spans 11/06/11 – 2/08/15. The ARX model is trained using data from the 2011-2012 and the 2012-2013 flu seasons.

*FNY Data*

FluNearYou (FNY) [15,16] compiles weekly data of ILI activity in the United States. They achieve this by conducting weekly, year-round, Internet-based surveys of voluntary participants who indicate whether they are healthy or have any of the following symptoms: fever, cough, sore throat, shortness of breath, chills/night sweats, fatigue, nausea/vomiting, diarrhea, body aches, headaches. FNY also collects data on the participant's location, vaccination status, gender, and age. We produced FNY ILI national estimates following the methodology introduced in Smolinski et al 2015 [16]. We used FNY data for weeks 10/24/11 to 02/21/15.

*GFT Data*

Google Flu Trends' weekly ILI national estimates are freely available through the Google Flu Trends website (www.google.org/flutrends). GFT data is the result of Google's proprietary algorithm that combines the volume of specific Google search queries to estimate the level of ILI activity in a given region [6, 28, 29]. We used GFT data for weeks 11/10/12 to 02/21/15, obtained from the http://www.google.org/flutrends/ website. This historical dataset was produced with the corresponding GFT engine active at the time the data was originally posted [29] (https://www.google.org/flutrends/about/how.html).

## 3. Methods

We chose three different machine learning algorithms: Stacked linear regression, Support Vector Machine regression, and AdaBoost with Decision Trees Regression, in order to optimally combine the five ILI estimates, produced independently with the five available data sources. We chose this set of machine learning algorithms since each one of them is known to have distinct strengths in combining information [30]. While the linearity assumption may be restrictive, we chose Stacked Linear Regression for simplicity. We chose Support Vector Machines (SVM) with radial basis function kernels because they map the input space to an infinite dimensional, nonlinear feature space, thus allowing more freedom on the functional relationship between the target and independent variables. Both Stacked Linear Regression and Support Vector Machines are global methods that apply the same rules to all of the data. We chose AdaBoost with Decision Trees because it has the power to learn local rules.

In the following paragraphs we describe the main features of each methodology.

*Stacked Linear Regression*
Stacked linear regression is a machine learning methodology commonly used in finance to combine weak predictors of stock prices [30, 31]. The goal of this methodology is not to identify which (so called) "weak predictor", $v_k(t)$, is the best one to predict the quantity $y(t)$ (in our case flu activity), but to linearly combine the information contained in all the "weak predictors" to obtain a more accurate and robust single predictor of a quantity $y(t)$. A multivariate approach is used to determine the best linear combination of weak predictors capable of producing the best prediction of the quantity $y(t)$ over a training period. Since the weak predictors are, by construction, highly correlated (indeed, each individual predictor was designed to minimize the square error between the predictions and flu activity), a way to discard redundant information is needed. Regularized approaches that penalize the size of the multiplying coefficients, $\alpha_k$, in the multivariate regression, such as Ridge or LASSO regularizations (L$_2$ and L$_1$, respectively), are good candidates to handle this. We chose LASSO regularization for our ensemble approach since we are interested in identifying models with the smallest number of independent variables ($v_k(t)$). Additionally, a non-negative constraint for each multiplicative coefficient $\alpha_k$ is imposed. This linear combination is then used to predict the value of $y(t)$ for values of $t$ outside of the training period.

*Support Vector Machine Regression*
Support Vector Machine (SVM) models [32] are similar to multivariate linear regression models with the important difference that non-linear functions can be chosen as the best relationship between the variables. This is achieved by introducing transformations (called kernels) that map the independent variables to higher dimensional feature spaces. The independent variables can even be mapped to an infinite dimensional feature space with the use of a radial basis function (RBF) kernel. SVM models are fitted by minimizing an epsilon-insensitive cost function where errors (between the predictions and the observed values) of magnitude less than epsilon are ignored in the cost function. This approach typically leads to better generalization of the chosen model on out-of-sample data. The SVM kernel type, margin width, and regularization hyper parameters were chosen via cross-validation on the training data.

*AdaBoost Regression with Decision Trees*
Decision Tree models are created by recursively splitting the input space, creating local models in each region of the input space. Decision trees, however, have been shown to be unstable as small changes in the data can lead to drastically different tree structures. Boosting methods, such as Adaptive Boosting (AdaBoost), are often employed to fix this problem. Adaptive Boosting (AdaBoost) regression [33] fits a sequence of weak learners (in this case decision trees) on sequentially reweighted versions of the training data. At each iteration, the weights are individually modified so that the training examples incorrectly predicted by the previous decision tree are given more importance when training the next decision tree. The final prediction is obtained by taking the weighted median of the predictions outputted by the ensemble of weak learners (AdaBoost.R2 algorithm: [33]).

*Independent variables*

In all of the aforementioned regression approaches the goal was to use all available information, in a given point in time, to produce accurate predictions of CDC's %ILI one, two, three, and four weeks ahead of the release of CDC reports, effectively predicting ILI three weeks into the future. At a given point in time, historical values up to two weeks prior to current date were available for all data sources (CDC, FNY, ATH, GT, GFT, and TWT). In addition real-time ILI estimates were available, with one-week lag, for ATH, GT, GFT, TWT. With this information, we produced predictions for every week starting on July 06, 2013 and up to February 21, 2015. For our first prediction, on the week of July 06, 2013, the first training set included 31 weeks worth of historical data from all data sources. For subsequent weeks, we dynamically increased the training set to include all available information at the given date, from all data sources.

*Baseline predictions*

As a reference, we produced ILI predictions using only historical CDC reported ILI. We achieved this via an autoregressive model with three weekly lagged components as independent variables (equation 1 in Paul et al 2014 [11]). We trained this model for the time period 11/06/11 – 2/08/15, and produced out-of-sample predictions for the four weekly time horizons during the time period of our study. We used the same procedure as the ARX model for Twitter, training on the 2011 – 2012 and 2012 – 2013 flu seasons, and producing predictions on the 2013 - 2014 and 2014 – 2015 flu seasons. These predictions were used to assess the added value provided by our digital disease detection systems' information.

*Evaluation metrics*

We report 5 evaluation metrics to compare the performance of the five independent predictors and the multiple ensemble methods: Pearson correlation, root mean squared error (RMSE), maximum absolute percent error (MAPE), Root Mean Square Percent error (RMSPE), and hit rate.

The definitions of all evaluation metrics are given below. Our notation is as follows: $y_i$ denotes the observed value of the CDC's ILI at time $t_i$, $x_i$ denotes the predicted value by any model at time $t_i$, $\bar{y}$ denotes the mean or average of the values $\{y_i\}$ and similarly $\bar{x}$ denotes the mean or average of the values $\{x_i\}$.

Pearson Correlation, a measure of the linear dependence between two variables during a time period $[t_1, t_n]$, is defined as:

$$r = \frac{\sum_{i=1}^{n}(y_i - \bar{y})(x_i - \bar{x})}{\sqrt{\sum_{i=1}^{n}(y_i - \bar{y})^2}\sqrt{\sum_{i=1}^{n}(x_i - \bar{x})^2}}$$

Root Mean Squared Error (RMSE), a measure of the difference between predicted and true values is defined as:

$$RMSE = \sqrt{\frac{1}{n}\sum_{i=1}^{n}(y_i - x_i)^2}$$

Root Mean Squared Percent Error (RMSPE), a measure of the percent difference between predicted and true values is defined as:

$$RMSPE = \sqrt{\frac{1}{n}\sum_{i=1}^{n}\left(\frac{y_i - x_i}{y_i}\right)^2} \times 100$$

Maximum Absolute Percent Error (MAPE), a measure of the magnitude of the maximum percent difference between predicted and true values, is defined as

$$MAPE = \left(\max_{i} \frac{|y_i - x_i|}{y_i}\right) \times 100$$

Hit Rate, a measure of how well the algorithm predicts the direction of change in the signal (independently of the magnitude of the change), is defined as:

$$Hit\ Rate = \frac{\sum_{i=2}^{n}(sign(y_i - y_{i-1}) == sign(x_i - x_{i-1}))}{n - 1} \times 100$$

where the symbol == denotes an if statement that returns the value 1, if the signs of predicted and observed changes are the same, and 0 otherwise.

These metrics were calculated for the time period: July 06, 2013 to February 21, 2015.

**Results**

*Real time estimates*
Table 1 presents the performance of the 5 *real-time* (nowcast) weak predictors as measured by each individual evaluation metric. This table is labeled "last week" since at a given point in time the *revised version* of all these estimates is only available on the Sunday of the reported week (or Monday of the subsequent week) and thus the information effectively predicts the %ILI of last week. For context, we included the metrics of three additional real-time predictions: (1) the *baseline* autoregressive predictions described in the previous section; (2) the CDC's Virology data, and (3) the best real-time ensemble method predictions, produced with a support vector machine (with RBF kernel).

As Table 1 shows, the *real-time* ensemble predictions outperform any individual weak predictor in all but one metric (the hit rate). A 0.989 Pearson correlation and an average error of about 0.176 %ILI (RMSE) make the ensemble approach a very accurate predictor. The ensemble predictions are very robust, as indicated by the size of the MAPE, which measures that the ensemble method is off-target, with respect to the *revised* CDC's ILI, by 23.6% in its worst occurrence in the whole time period (with comparable performance to LASSO's 20.2% MAPE). This error is smaller than two thirds of the smallest MAPE of any of the individual weak predictors. In terms of hit rate, which reflects the ability of the method to predict the upward or downward tendency of the CDC's ILI (in addition to the Pearson correlation and independently of producing an accurate point estimate, as captured by RMSE), athenaheath data (ATH) offers the best results.

Furthermore, Table 1 quantitatively shows the added value of using real-time digital disease detection information over a simple historical autoregressive approach. This can be seen by the improvement of the Pearson correlation from 0.930 to 0.989, the near three-fold reduction on the RMSE, and the maximum absolute error cut in half.

The top panel of Figure 1 graphically shows the *revised* CDC's ILI along with the predictions of: the 5 data sources, the baseline, and the best ensemble approach (SVM RBF), as a function of time. The errors for each predictor are displayed in the bottom panel of Figure 1. The real-time estimates produced with our ensemble method are capable of predicting the timing and magnitude of the two peaks of the 2014-2015 season exactly, whereas they predict the peak of the 2013-2014 season with a one-week lag. Overall predictions track very accurately the CDC's revised %ILI.

*Forecasts*
Since none of the five weak predictors produce predictions into the future (forecasts), we do not have the equivalent of Table 1 for the three forecast time horizons (labeled "this week", "next week", and "in two weeks"). Table 2 presents the performance of 4 different machine learning ensemble approaches and the baseline autoregressive predictions for the four time horizons. Figure 2 shows these results graphically. Ensemble predictions produced with the AdaBoost method show the best accuracy (lowest RMSE) and robustness (lowest MAPE), for the three forecast time horizons. Correlation is also highest with AdaBoost in all three horizons. While the hit rate seems to be highest for different methods in different time horizons, Adaboost has an overall best performance as observed in Figure 3. We highlight the fact that our ensemble predictions one week into the future, labeled "this week", have comparable accuracy to *real-time* GFT predictions, as measured by RMSE.

As shown in Table 2, our ensemble approach produces better results than the baseline AR3 autoregressive model in all similarity metrics and all time horizons. This fact shows quantitatively the value of using social media and crowd-sourced data in improving influenza predictions in future %ILI predictions. Specifically, the average error (RMSE) of our ensemble predictions nearly halves the errors of the autoregressive predictions in all time horizons. Pearson correlations of our ensemble approach predictions improve their autoregressive counterparts, from 0.845 to 0.960, in the one week forecast; from 0.759 to 0.927, in the two-week forecast, and from 0.683 to 0.904, in the three week forecast. Note also that our forecast estimates in all time horizons (up to four weeks ahead of the release of CDC's reports) show at least comparable accuracy to "*real-time*" estimates obtained with a purely autoregressive model.

The ability of the ensemble approach forecasts to capture the timing and magnitude of the peaks in the flu seasons decays as the time horizon increases, as observed in Figure 2. Indeed, one-week forecasts predict the 2013-2014 peak with a one-week lag and with a percent error of about 10%, and they predict the two 2014-2015 peaks with a one-week lag and with percentage errors less than 2%. The two-week forecasts capture the 2013-2014 peak with a one-week lag and show percentage errors of about 10%, and they predict the two 2014-2015 peaks with a two-week lag and percentage errors up to 20%. Finally, the three-week forecasts capture the 2013-2014 peak with a two week lag and show percentage errors of about 20%, and they predict the two 2014-2015 peaks with a two-three week lag and with percentage errors up to 25-30%.

**Discussion**

Our results show that our real-time ensemble predictions outperform every real-time flu predictor constructed independently with each data source. This fact suggests that combining information from multiple independent flu predictors is advantageous over simply choosing the best performing predictor. This is the case not only for real-time predictions but also for the one, two and three week forecasts presented.

Specifically, we show that our methodology can produce predictions one week ahead of GFT's real-time estimates with comparable accuracy. We also show that our ensemble forecasts (up to three weeks into the future) always improve predictions produced with a baseline autoregressive model, thus proving quantitatively the added value of incorporating search and social media data in our flu prediction models.

It is interesting to highlight that the correlation and RMSE of the ensemble approach real-time predictions (Corr: 0.989 and RMSE: 0.176) are similar to the differences between *revised* and *unrevised* CDC reports (Corr: 0.993 and RMSE: 0.162). This means that our real-time ensemble model is as accurate a predictor of the *revised* CDC's ILI estimates as the unrevised CDC data is. Thus, it is possible that we may be reaching the limit of what is possible, in terms of producing an accurate predictor of *revised* CDC's ILI.

Our ensemble estimates correlate better with CDC's ILI than CDC's Virology data (which measures lab-confirmed cases of influenza) does with CDC's ILI. This suggests that our (search and social media) data sources, when combined appropriately, track closely people showing symptoms and not necessarily those that are confirmed with influenza. It is important to mention that CDC's Virology data (http://www.cdc.gov/vaccines/pubs/surv-manual/chpt06-influenza.html) is not necessarily considered to be a good predictor of ILI and tends to be even more lagged than CDC's ILI due to the slowness of laboratory testing [34,35,36].

Doubts have emerged regarding the value of digital disease detection methods as a consequence of the multiple discrepancies between GFT's predictions and the observed CDC's ILI estimates [18,19, 20, 21, 22, 24, 28, 29]. We highlight the fact that even when one of the independent predictors produces unreliable estimates, our ensemble estimates are robust and accurate. This is observed specifically during the 2014-2015 flu season when ATH and GFT overestimated the flu season peak magnitude by more that 30% and approximately 15%, respectively, and the real-time ensemble approach estimates were right on target.

An additional attribute of our approach is that even if the ground truth (now CDC reports) were chosen to come from a different (and potentially more appropriate) source, our methodology would seamlessly adapt to predicting any target signal.

While the results presented here are for influenza-like illnesses at the national level within the US, our approach shows promise to be easily extended to accurately track not only influenza in other countries where multiple data sources may be available [37,38] but also other infectious diseases. Indeed, infectious diseases such as Dengue [39, 40, 41] or Malaria [42], for which multiple surveillance methods are in place would benefit from combining information in a similar way to the one proposed here. Moreover, disease surveillance data at finer spatial resolutions tend to be scarcer and often unreliable [43], and thus, approaches like ours may help produce more accurate and robust disease incidence estimates, at higher spatial resolutions, by drawing data from multiple sources.

*Limitations*
Using weekly information from reports published by the CDC as our gold standard for national flu activity may not necessarily be ideal. Indeed, two data sources considered in this study, athenahealth and Flu Near You, aim at tracking the percentage of the general population with ILI symptoms independently. While athenahealth can be thought of as a subsample of the CDC-reported %ILI (since it calculates the %ILI in a similar fashion to the CDC, except with the information from those patients seeking medical attention in facilities managed by athenahealth), Flu Near You aims at providing an estimate of flu activity from a potentially distinct population (people willing to report their health status in weekly surveys via a mobile phone app). Interestingly, while the sectors of the population sampled by the CDC and FNY maybe distinct (they may overlap when people report their symptoms using the FNY app and they seek medical attention), Figure 1 and a recent study [16] show that their ILI estimates track one another quite well (Pearson correlation of .948) suggesting that both FNY and CDC datasets may be good proxies of ILI activity in the population. Finally, the best ensemble methodology may change for future flu seasons, and thus, continuous monitoring of the multiple methodologies' performances should be conducted as new predictions are produced.

**Conclusion**
We presented a methodology that optimally combines the information from multiple real-time flu predictors to produce more accurate and robust real-time flu predictions than any other existing system. Moreover, our ensemble approach is capable of using real-time and historical information to accurately forecast flu estimates one, two, and three weeks into the future.

**Tables**

|  | CORR | RMSE (%ILI) | RMSPE (%) | MAPE (%) | Hit Rate |
|---|---|---|---|---|---|
| FNY | 0.948 | 0.385 | 15.9 | 39.3 | 65.9 |
| ATH | 0.977 | 0.351 | 14.1 | 36.7 | **77.7** |
| GT | 0.978 | 0.245 | 13.3 | 42.9 | 65.9 |
| GFT | 0.980 | 0.333 | 12.3 | 35.3 | 75.3 |
| TWT | 0.937 | 0.414 | 15.1 | 50.1 | 62.4 |
| CDC Baseline | 0.930 | 0.501 | 18.2 | 46.7 | 68.2 |
| CDC Virology | 0.923 | - | - | - | 69.4 |
| SVM (RBF) | **0.989** | **0.176** | **8.27** | **23.6** | 69.4 |

**Table 1.** Similarity metrics between CDC's ILI and the 5 weak predictors: Flu Near You, athenahealth, Google Trends, Google Flu Trends, and Twitter, for the time period Aug 2013 - Feb 2015. For reference, an autoregressive model (AR3) was utilized as a baseline. Pearson correlation and Hit rate for CDC's Virology data are shown. The best performing model per metric is bold faced.

|            | CORR  | RMSE (%ILI) | RMSPE (%) | MAPE (%) | Hit Rate |
|------------|-------|-------------|-----------|----------|----------|
| LASSO      | 0.986 | 0.213       | 9.36      | **20.2** | **75.3** |
| SVM (RBF)  | **0.989** | **0.176** | **8.27** | 23.6    | 69.4     |
| SVM (Linear) | 0.988 | 0.217     | 8.39      | 23.5     | 74.1     |
| AdaBoost   | 0.977 | 0.251       | 10.4      | 34.0     | 63.5     |
| CDC Baseline | 0.930 | 0.501     | 18.2      | 46.7     | 68.2     |

Table 1: Last Week

|            | CORR  | RMSE (%ILI) | RMSPE (%) | MAPE (%) | Hit Rate |
|------------|-------|-------------|-----------|----------|----------|
| LASSO      | 0.921 | 0.477       | 19.4      | 52.1     | 71.8     |
| SVM (RBF)  | 0.955 | 0.352       | **12.8**  | 35.9     | **75.3** |
| SVM (Linear) | 0.926 | 0.469     | 14.7      | 52.4     | **75.3** |
| AdaBoost   | **0.960** | **0.334** | 13.1    | **34.1** | 62.4     |
| CDC Baseline | 0.845 | 0.709     | 24.2      | 51.5     | 70.6     |

Table 2: This Week

|            | CORR  | RMSE (%ILI) | RMSPE (%) | MAPE (%) | Hit Rate |
|------------|-------|-------------|-----------|----------|----------|
| LASSO      | 0.788 | 0.766       | 29.4      | 106      | 61.2     |
| SVM (RBF)  | 0.894 | 0.527       | 26.9      | 137      | **65.9** |
| SVM (Linear) | 0.799 | 0.764     | 26.8      | 124      | 49.4     |
| AdaBoost   | **0.927** | **0.446** | **16.1** | **41.0** | 51.8   |
| CDC Baseline | 0.759 | 0.863     | 29.4      | 66.8     | 62.4     |

Table 3: Next Week

|            | CORR  | RMSE (%ILI) | RMSPE (%) | MAPE (%) | Hit Rate |
|------------|-------|-------------|-----------|----------|----------|
| LASSO      | 0.713 | 0.857       | 36.5      | 92.4     | 60.0     |
| SVM (RBF)  | 0.902 | 0.507       | 27.4      | 131      | 54.1     |
| SVM (Linear) | 0.697 | 0.896     | 28.1      | 86.6     | 58.8     |
| AdaBoost   | **0.904** | **0.503** | **20.3** | **47.7** | 52.9   |
| CDC Baseline | 0.683 | 0.977     | 33.7      | 75.5     | **62.4** |

Table 4: In Two Weeks

**Table 2.** Similarity metrics between CDC's ILI and 4 machine learning ensemble methods for last week (top), this week (second), next week (third), and two weeks from now (bottom), for the time period Aug 2013 - Feb 2015. For reference, an autoregressive model (AR3) was utilized as a baseline. The best performing model per metric is bold faced.

## Figures

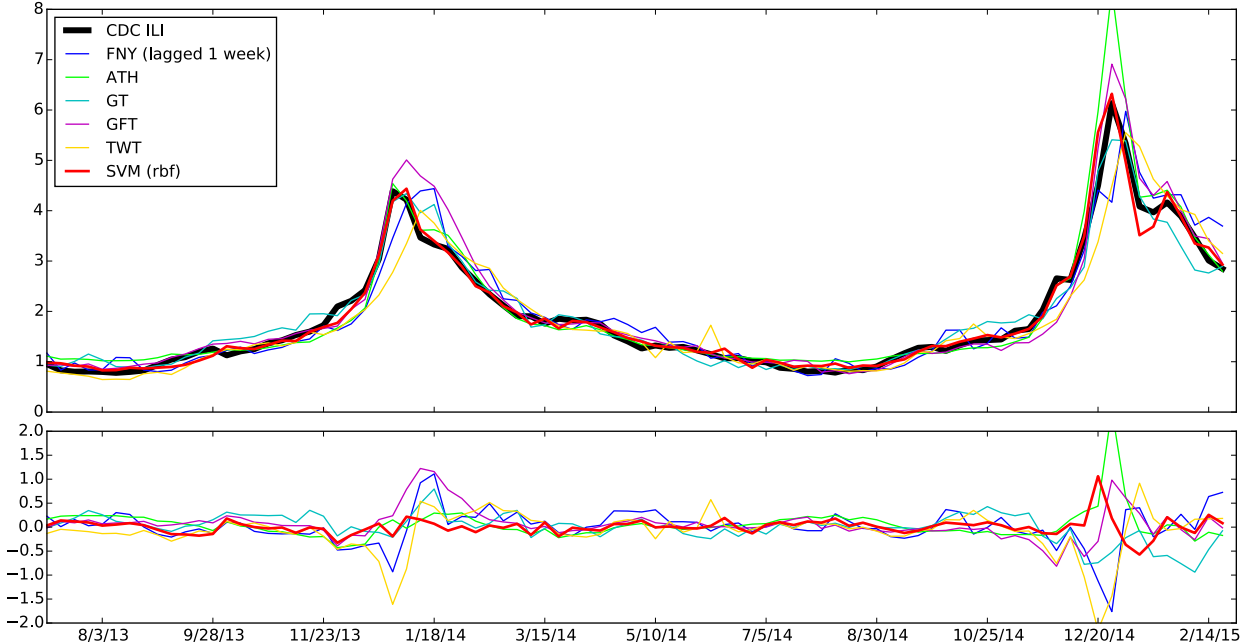

**Figure 1.** The CDC's %ILI (Influenza like illnesses), the performance of the 5 available predictors, the baseline predictions, and the performance of the best ensemble method for last week's predictions are displayed as a function of time (top). The errors associated with each weak predictor and the ensemble approach are shown (bottom).

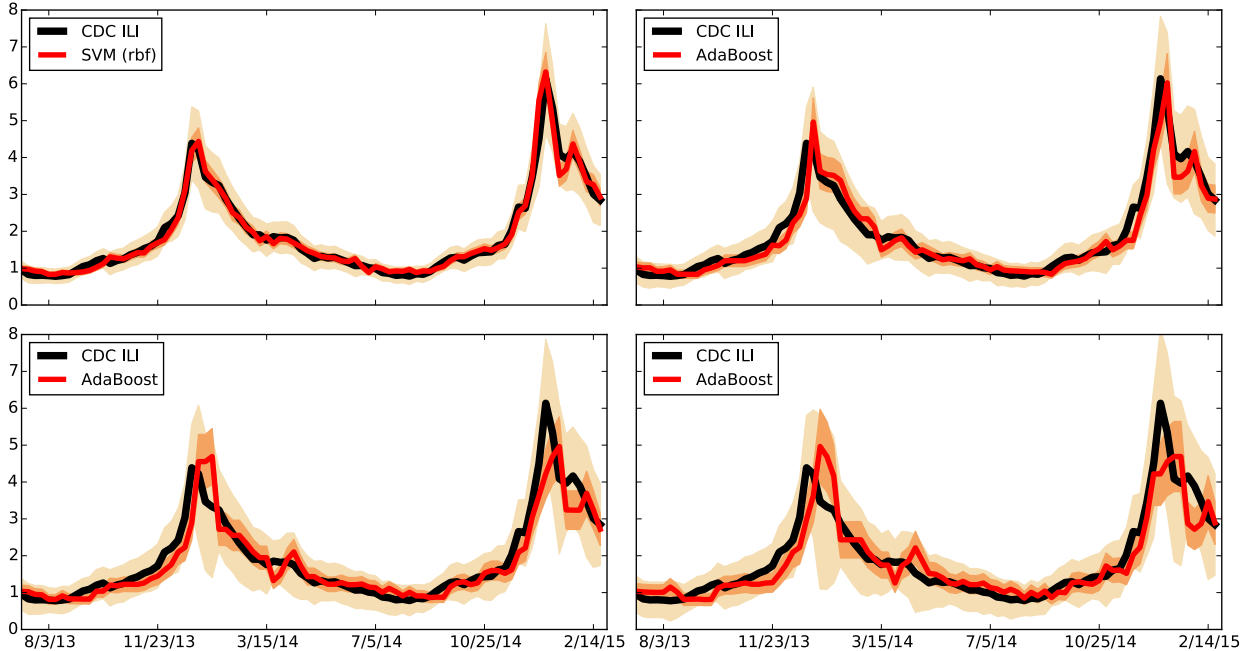

**Figure 2.** The best performing ensemble approach is shown in red side by side to the CDC's % ILI for all time horizons: last week (top left), current week (top right), next week (bottom left), and two weeks from current (bottom right). The dark error bars correspond to the relative root mean squared error (RRMSE) and the light error bars correspond to the relative maximum absolute error.

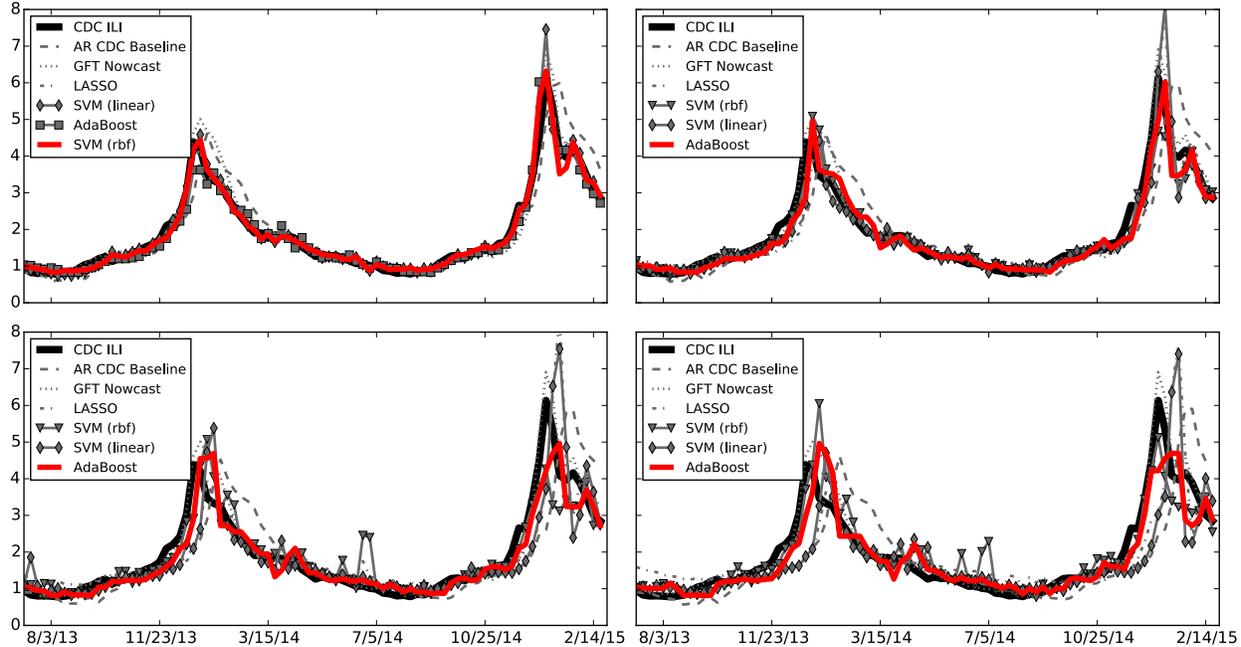

**Figure 3.** The CDC's %ILI (Influenza like illnesses) and the performance of multiple machine learning ensemble approaches that combine the 5 weak predictors to produce a single estimate are displayed for comparison for the four time horizons: last week (top left), current week (top right), next week (bottom left), and two weeks from current (bottom right). The red curve displays the performance of the best method for a given time horizon. As expected, the accuracy and robustness of the predictions decrease as the time horizon increases.

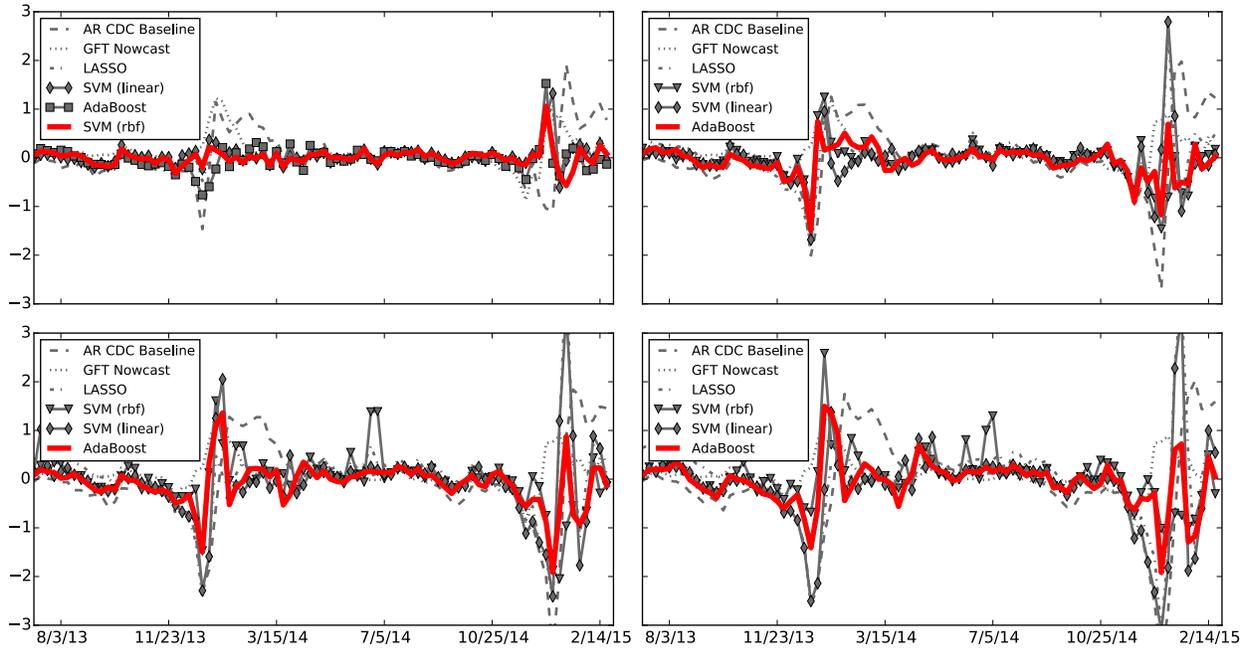

**Figure 4.** Errors associated with each ensemble approach are displayed for all time horizons: last week (top left), current week (top right), next week (bottom left), and two weeks from current (bottom right).